\documentclass[english,aps,pra,twocolumn,groupedaddress,floatfix]{revtex4}

\usepackage{amsmath}
\usepackage{amssymb}
\usepackage{amsfonts}
\usepackage{bbm}
\usepackage{epsfig}
\usepackage{grffile}
\usepackage{babel}

\usepackage{color}

\newcommand{\id}{\mathbbm{1}}
\newcommand{\RR}{{\mathbbm{R}}}
\newcommand{\Id}{{\operatorname{Id}}}
\newcommand{\Tr}{\operatorname{Tr}}
\newcommand{\Span}{\operatorname{span}}
\newcommand{\bra}{\langle}
\newcommand{\ket}{\rangle}
\newcommand{\T}{{\!\top}}
\renewcommand{\vec}[1]{{\boldsymbol{#1}}}
\newcommand{\mc}[1]{\mathcal{#1}}
\newcommand{\pdag}{{\phantom{\dag}}}

\newcommand{\NN}{\mathbb{N}}

\newcommand{\B}{\mc{B}}
\newcommand{\Csq}{{\mc{C}}}
\newcommand{\D}{\mc{S}}
\newcommand{\F}{\mc{F}}
\newcommand{\G}{\mc{G}}
\newcommand{\R}{\mc{R}}
\newcommand{\none}{\text{\o}}

\newcommand{\dm}{{\hat{\rho}}}
\newcommand{\A}{\hat{a}}
\renewcommand{\S}{\hat{S}}
\newcommand{\sig}{{\hat{\sigma}}}
\newcommand{\hC}{\hat{C}}
\newcommand{\hH}{\hat{H}}

\newcommand{\tmat}[1] {\begin{bmatrix}#1\end{bmatrix}}

\newcommand{\blue}[1] {\textcolor{blue}{#1}}

\usepackage{hyperref}

\newcommand{\fu}{Dahlem Center for Complex Quantum Systems, Freie Universit{\"a}t Berlin, 14195 Berlin, Germany}
\newcommand{\potsdam}{Institute for Physics and Astronomy, University of Potsdam, 14476 Potsdam, Germany}

\newcommand{\Title} {Solving condensed-matter ground-state problems by semidefinite relaxations}
\newcommand{\Authors}
{
\author{Thomas Barthel}
\author{Robert H\"{u}bener}
\affiliation{\fu}
\affiliation{\potsdam}
}
\newcommand{\Date} {February 12, 2012}

\begin{document}

\title{\Title}
\Authors

\date{\Date}

\begin{abstract}
We present a new generic approach to the condensed-matter ground-state problem which is complementary to variational techniques and works directly in the thermodynamic limit. Relaxing the ground-state problem, we obtain semidefinite programs (SDP). These can be solved efficiently, yielding strict lower bounds to the ground-state energy and approximations to the few-particle Green's functions. As the method is applicable for all particle statistics, it represents in particular a novel route for the study of strongly correlated fermionic and frustrated spin systems in $D>1$ spatial dimensions. It is demonstrated for the XXZ model and the Hubbard model of spinless fermions. The results are compared against exact solutions, quantum Monte Carlo, and Anderson bounds, showing the competitiveness of the SDP method.
\end{abstract}


\maketitle

\section{Introduction}
Prominent simulation techniques for condensed-matter systems are sampling algorithms like \emph{quantum Monte Carlo} (QMC) \cite{Suzuki1977-58,Prokofev1996-64,Syljuasen2002-66,Alet2005-71} and variational algorithms like the \emph{density-matrix renormalization group} (DMRG) \cite{White1992-11,Schollwoeck2005}, other \emph{tensor-network-state} (TNS) approaches \cite{Niggemann1997-104,Nishino2000-575,Verstraete2004-7,Vidal-2005-12}, or \emph{variational Monte Carlo} \cite{McMillan1965-138,Ceperley1977-16,Mezzacapo2009-11,Changlani2009-80}. For a number of interesting classes of systems, like frustrated or fermionic systems in $D>1$ spatial dimensions, the powerful QMC technique is inefficient due to the \emph{sign problem} \cite{Hirsch1982-26,Takasu1986-75,Troyer2005}. Such systems are then often studied with variational techniques by minimizing the energy within a certain class of states as, e.g., TNS of a certain structure. The energy expectation value of the obtained state is necessarily an upper bound to the exact ground-state energy.

In this article, a complementary approach is presented. By relaxations of the ground-state problem we obtain \emph{semidefinite programs} (SDP) \cite{Vandenberghe1996-38,Alizadeh1995-5}, which can be solved efficiently on classical computers. This yields lower bounds to the ground-state energy and corresponding approximations to few-particle Green's functions. The obtained Green's functions allow, e.g., for the study of phase diagrams. As the presented SDP method works irrespective of the particle statistics, it provides in particular a novel route for the study of strongly correlated fermionic and frustrated spin systems for $D>1$.
The method can also be used to judge the quality of variational algorithms in situations where exact or QMC results are not available for comparison. This is especially important for variational Monte Carlo methods \cite{Mezzacapo2009-11,Changlani2009-80} and the recently developed variational TNS techniques for fermions in $D>1$ \cite{Kraus2009_04,Corboz2009_04,Pineda2009_05,Barthel2009-80,Corboz2009-80}.

The idea is to specify the system by its equal-time $k$-point Green's functions $\G^{(k)}$. For systems of fermions, bosons, or hard-core bosons (being equivalent to spins-$1/2$) in a normalized state $\dm$, they are defined as the correlation functions
\begin{equation}\label{eq:GFdef}
	\G^{(k)}_{\vec{i},\vec{j}}:=\Tr(\dm\,\A_{i_1}^\pdag\dots\A_{i_m}^\pdag \A_{j_n}^\dag\dots\A_{j_1}^\dag ),\quad k=m+n,
\end{equation}
for ladder operators $\A_i$ and some single-particle basis states $|i\ket$; \cite{Negele1988}.
The energy expectation value $E=\Tr\dm\hH$ is a function of the (few-particle) Green's functions, i.e., $E=E(\G)$.
The exact ground-state energy is obtained by minimization among all \emph{representable} Green's functions $\G$, i.e., those for which a density operator $\dm$ exists, such that Eq.~\eqref{eq:GFdef} is obeyed. However, to determine whether a given Green's function $G$ is representable turns out to be a computationally hard problem, the famous \emph{$N$-representability problem} \cite{Tredgold1957-105,Coulson1960-32,Coleman1963-35,Deza1994-55}, which is QMA-complete \cite{Liu2007-98}. Nevertheless, an efficient minimization of $E$ is possible if we relax the constraints on the Green's function, then, yielding not the exact ground-state energy but a lower bound, and not the exact ground-state Green's function but an approximation. 
Note that minimizing $E(G)$ without any constraints is doomed to fail, as the energy is linear in $G$.
Manageable constraints on $G$ can be constructed by imposing the positivity of the expectation values of certain positive-definite observables, an example being the particle density operator $\A_i^\dag\A_i^\pdag$.
Such an approach for fermionic $G^{(1)}$ and $G^{(2)}$ has been successfully applied to finite systems in quantum chemistry; see, e.g., Ref.~\cite{Mazziotti2006-39} and references therein.

Here, we present a systematic method for the construction and solution of relaxed ground-state problems for condensed-matter systems. As in the approach known from quantum chemistry, our constraints enforce positive expectation values for operators of the form $\hC^\dag\hC$ with respect to the Green's functions $G$. More generally than before, we (a) choose the \emph{constraint operators} $\hC$ to be arbitrary polynomials of degree $\leq K\in\NN$ in the ladder operators, which (b) act on suitably chosen subsets of the lattice. This makes it possible to address large systems and to control the precision and the computation cost. We (c) exploit the translation invariance of the condensed-matter systems to (d) work effectively with an infinite number of degrees of freedom and describe the systems directly in the thermodynamic limit. 
The method yields coupled constraints for  $G^{(1)},\dotsc,G^{(2K)}$ such that the energy optimization problem attains the form of an SDP. On basis of the bipolar theorem \cite{Rockafellar1970}, we further elucidate the mathematical background of the method and further possible reductions to the number of constraints. 

An alternative method for calculating lower bounds to the ground-state energy is due to Anderson. \emph{Anderson bounds} \cite{Anderson1951-83,Wittmann1993-48} are obtained by splitting the Hamiltonian $\hH$ into a sum of subsystem Hamiltonians $\hH_m$ that are accessible by exact diagonalization. The Anderson bound for the ground-state energy $E^0$ is then given by the sum of the ground-state energies $E^0_m$ of the $\hH_m$,
\begin{equation}\label{AndersonBound}\textstyle
	\hH = \sum_m \hH_m\quad \Rightarrow\quad
	E^0 \geq \sum_m E^0_m.
\end{equation}
The computation cost for this bound scales exponentially in the sizes of the spatial supports of the operators $\hH_m$; see Appx.~\ref{sec:Anderson}. A generalization of this approach to finite temperatures is presented in Ref.\ \cite{Poulin2011-106}.

In contrast, the computation cost for the SDP method scales only polynomially in the support of the constraint operators.
In the prominent systems that we studied with moderate computer resources, the SDP method outperforms the Anderson bound substantially. The SDP method has the additional advantage of giving access to the Green's functions, which can be used to study phase diagrams, etc..
Furthermore, it is possible to choose different spatial supports of the constraint operators $\hC$ depending on their degree in the ladder operators. The optimal choice for those spatial supports depends on the position in the phase diagram.

\section{Ground-state problem}\label{sec:Problem}
The following description applies to lattice systems of
fermions, bosons, and hard-core bosons -- each corresponding to a certain algebra for the ladder operators $\{\A_i,\A_i^\dag\}$ \cite{Negele1988}. Spins-$1/2$  can be treated by mapping them to hard-core bosons via the identification $\A_i^\pdag=\S_i^{-}$, $\A_i^\dag=\S_i^{+}$, and $\S^z_i=\A_i^\dag\A_i^\pdag-\frac{1}{2}$. The generalization to higher spins is straightforward.
For each subset $\Omega$ of the single-particle modes $\{|i\ket\}$, let $\mc{A}^k_\Omega$ denote the operator basis of normal-ordered monomials of degree $k$ in the ladder operators for subsystem $\Omega$.
\begin{equation}\label{eq:opBasis}
	\mc{A}^k_\Omega:=\{ 
	\A_{i_1}^\pdag\dots\A_{i_m}^\pdag \A_{i_{m+1}}^\dag\dots\A_{i_k}^\dag\,|\,
	0\leq m\leq k,\, i_\ell\in \Omega\}
\end{equation}
Each density operator $\dm$ corresponds to a representable Green's function $\G$, with its $k$-point component given by
\begin{equation}
	\G^{(k)}_\sig := \Tr\dm\sig
	\quad\text{for}\quad \sig\in \mc{A}^k_\Omega.
\end{equation}
For the moment, let us choose $\Omega$ to be the full system. Every linear operator $\hat B$ on the Hilbert space can be expanded in the basis $\mc{A}:=\bigcup_k\mc{A}^k$ as $\hat B=\sum_{\sig\in\mc{A}} B_\sig \sig$. Its expectation value with respect to a state $\dm$ is then
\begin{equation}
	\Tr\dm \hat B = \sum_{\sig\in \mc{A}} \G_\sig B_\sig  =: \G[\hat B],
\end{equation}
where $\G$ is the Green's function of $\dm$. In this sense, Green's functions are linear functionals on the operators.

For a Hamiltonian $\hH$, the ground-state problem reads
\begin{equation}\label{eq:gsp}
	E^0=\min_{\dm\in\D} \Tr \dm\hH
	 = \min_{\G\in \R} \G[\hH],
\end{equation}
where $\D$ denotes the set of all density operators and $\R$ denotes the set of all representable Green's functions, 
\begin{align}
	\R:&=\{\G\,|\, \exists\,\dm\in\D:\, \G_\sig=\Tr\dm\sig \,\,\,\forall_{\sig\in \mc{A}} \} \nonumber\\\label{eq:representableG}
	 &= \{\G\,|\,\G[\Id]=1,\, \G_{\sig^\dag}=\G^*_\sig,\, \G[\hat B]\geq 0\,\,\,\forall_{\hat B\succeq 0}\}.
\end{align}
This equality follows from the fact that the only constraints on a valid density operator $\dm$ are $\Tr\dm=1$, $\dm=\dm^\dag$, and its positivity $\dm\succeq 0$, which is equivalent to requiring $\Tr\dm\hat B\geq 0$ for all positive-semidefinite operators $\hat B\succeq 0$.

Variational methods proceed from Eq.~\eqref{eq:gsp} by choosing some accessible subset of $\D$. Each variational state from such a subset yields an upper bound to the ground-state energy $E^0$. In contrast, for the SDP method, described in the following, one chooses an accessible superset $\F$ of the set $\R$ of representable Green's functions, i.e., relaxes the constraints. Minimizing the energy in such a superset yields a lower bound to $E^0$. A decisive feature of the SDP method is that the minimum energies for the chosen supersets $\F$ can be found certifiably; see Appx.~\ref{sec:sdp}.

\section{SDP method}\label{sec:Method}
Solving the ground-state problem \eqref{eq:gsp} in general is known to be a computationally hard problem; it is QMA-complete \cite{Kempe2006-35}. Similarly, determining whether given Green's functions are representable, i.e., elements of $\R$ in Eq.~\eqref{eq:representableG}, is also a QMA-complete problem \cite{Liu2007-98}.

 A straightforward way to relax the -- apparently too demanding -- constraints represented by $\R$ is to require the Green's functional $G[\hat B]$ to be non-negative, not for all $\hat B\succeq 0$ but only for operators $\hat B$ of the form $\hat B=\hC^\dag \hC$ with constraint operators $\hC$ from some suitable set $\Csq$. Minimizing the energy $G[\hH]$ with respect to Green's functions $G$ from the set
\begin{equation}\label{eq:GSetRelaxed}
	\F_\Csq := \{ G\,|\, G[\Id]=1,\, G_{\sig^\dag}=G^*_\sig,\, G[\hC^\dag\hC]\geq 0\,\,\,\forall_{\hC\in \Csq}\},
\end{equation}
yields a lower bound to the ground-state energy \eqref{eq:gsp}.
\begin{equation}\label{eq:gspRelaxed}
	E^0 \geq \min_{G\in \F_\Csq} G[\hH] \equiv \min_{G\in \F_\Csq}\sum_{\sig} G_\sig H_\sig,
\end{equation}
as $\R\subset\F_\Csq$.
Imposing more and more constraints by enlarging the operator set $\Csq$, the bound approaches $E^0$, and the optimal $G$ approaches the ground-state Green's function. 
Determining the optimum in Eq.~\eqref{eq:gspRelaxed} is a semidefinite programming problem for the variables $G_\sig$, as $ G[\hH]\equiv \sum_{\sig} G_\sig H_\sig$ is linear in $G$, and the constraints $G[\hC^\dag\hC]\geq 0$ $\forall\,\hC\in \Csq$ can be written in the form
\begin{equation}\label{eq:SDPconstraint}
	\sum_\sig G_\sig M_\sig \succeq 0,
\end{equation}
where the Hermitian matrices $M_\sig$ are completely determined by the underlying algebra of the ladder operators $\A_i$ and the choice for the constraint operator set $\Csq$. Eqs.~\eqref{eq:gspRelaxed} and \eqref{eq:SDPconstraint} correspond to a standard form for an SDP \cite{Vandenberghe1996-38,Alizadeh1995-5}; see also Appx.~\ref{sec:sdp}. Eq.~\eqref{eq:SDPconstraint} results from expanding the constraint operators in the basis $\mc{A}$, $\hC =\sum_\sig C_\sig\sig$. This yields the constraints in the form $\sum_{\sig'\sig''}C_{\sig'}^*G[(\sig')^\dag\sig''] C_{\sig''}\geq 0$ $\forall_\vec{C}$. Expanding the operators $(\sig')^\dag\sig''$ in the operator basis $\mc{A}$ (by bringing them into normal-ordered form) and using $G[\sig]\equiv G_{\sig}$ yields the matrices $[M_\sig]_{\sig',\sig''}$ and the constraints
\begin{equation*}
	\sum_{\sig \sig' \sig''} G_\sig C^*_{\sig'} [M_\sig]_{\sig',\sig''} C_{\sig''} = \sum_\sig G_\sig \vec{C}^\dag M_\sig \vec{C}\geq 0\quad  \forall_\vec{C}
\end{equation*}
from which Eq.~\eqref{eq:SDPconstraint} follows.

\section{Thermodynamic limit and constraint operators}\label{sec:tdLimit}
Let us now turn to the specific case of condensed-matter systems in the thermodynamic limit. Let the Hamiltonian be translation-invariant $\hat H=\sum_\vec{r} \mc{T}_\vec{r}(\hat{h})$ with finite-range interaction terms $\hat{h}$ and the lattice translation operator $\mc{T}_\vec{r}$. We denote the spatial support of the $\kappa$-point terms in $\hat{h}$ by $\Lambda_\kappa$. For a particle-number-conserving $D$-dimensional model with 2-body nearest-neighbor interactions, this means for example $|\Lambda_2|=|\Lambda_4|=D+1$, and $\Lambda_\kappa=\emptyset$ $\forall_{\kappa\neq 2,4}$. Due to the translation invariance of $\hat H$, we can restrict ourselves to translation-invariant density matrices and Green's functions.
A constraint operator set can be constructed by choosing, for each operator degree $k$, a subsystem $\Omega_k$ of the full lattice such that
\begin{equation}\label{eq:subsystems}
	\Omega_{k'}\subset\Omega_k\quad\forall_{k'>k}
	\quad\text{and} \quad
	\Lambda_{\kappa}\subset\Omega_{\lceil\kappa/2\rceil}\quad\forall_{\kappa}.
\end{equation}
Every such choice of subsystems and the corresponding set of constraint operators
\begin{equation}\label{eq:constraintOp}
	\Csq_{\vec{\Omega}}:= \Span \mc{A}_\vec{\Omega}\quad\text{with}\quad
	\mc{A}_\vec{\Omega}:=\bigcup_k \mc{A}^k_{\Omega_k}
\end{equation}
defines with Eq.~\eqref{eq:GSetRelaxed} a set $\F_{\Csq_\vec{\Omega}}\supset\R$ of Green's functions. The number of Green's function elements $G_\sig$ occurring as degrees of freedom in the SDP is then given by the size $|\mc{A}_\vec{\Omega}|$ of the operator basis.
It depends on the particle statistics: For a given subsystem $\Omega$ of a bosonic system, $|\mc{A}^k_\Omega|$ grows exponentially in the operator degree $k$. For fermions, hard-core bosons, and spins-$1/2$, $|\mc{A}^k_\Omega|$ initially grows exponentially but drops to zero for $k>2|\Omega|$.
We always choose some $K$ so that $\Omega_k=\emptyset$ $\forall_{k>K}$.
Enlarging the subsystems $\Omega_k$, systematically improves the solution of Eq.~\eqref{eq:gspRelaxed} and increases the computation cost polynomially.
For given model and computer resources, the optimal choice for the subsystems $\Omega_k$ depends on the position in the phase diagram.

\section{Symmetries}\label{sec:symmetries}
Hamiltonian symmetries, like translation or rotation invariance, imply that several Green's function elements $G_\sig$ can be chosen to be identical (e.g., $G_{\A^\pdag_\vec{x}\A^\dag_\vec{y}}\equiv G_{\A^\pdag_\vec{0}\A^\dag_{\vec{y}-\vec{x}}}$ $\forall_{\vec{x}\vec{y}}$) and it is sufficient to use in the SDP only one representative for each of the corresponding equivalence classes. Further, several Green's function elements can be zero (e.g., $G_{\A_\vec{x}\A_\vec{y}}=0$ for particle-number-conserving models). A corresponding block structure in $\sum_\sig G_\sig M_\sig$ can be exploited to further reduce the computation cost.

\section{Exemplary applications}
We demonstrate the capabilities of our SDP approach with three example systems. The first two are chosen such that one can compare against high-precision data from other methods. The third model is used to demonstrate that the SDP method is applicable to models that can not be treated satisfactorily by other means.
\subsection{XXZ chain}
\begin{figure}
	\includegraphics[width=\linewidth]{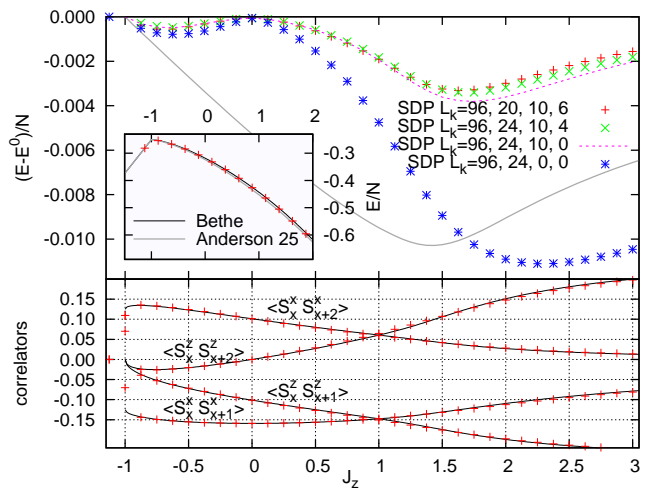}
\caption{\label{fig:XXZ1D} Lower bounds $E$ to the ground-state energy and approximations to correlators for the XXZ chain \eqref{eq:HamXXZ}. The Anderson bound \eqref{AndersonBound} was calculated with clusters of 25 sites. The subsystems $\Omega_{k}$ for the SDP method \eqref{eq:constraintOp} are chosen to be clusters of contiguous sites with sizes $L_1,L_2,L_3$, and $L_4$ for constraint operators of polynomial degree 1, 2, 3, and 4, respectively, as specified in the legend ($K=4$). The Bethe ansatz yields the exact ground-state energy $E^0$ and short-range correlators \cite{Cloizeaux1966-7,Kato2003-36}.}
\end{figure}
Let us first address the spin-$1/2$ XXZ model 
\begin{equation}\label{eq:HamXXZ}
	 \hat H = \sum_{\bra i,j\ket}(\S^x_i\S^x_j+\S^y_i\S^y_j+J_z \S^z_i\S^z_j)
\end{equation} 
in one spatial dimension (1D).
For $J_z<-1$, the model is in a gapped ferromagnetic phase with a fully polarized ground state. In the region $-1\leq J_z\leq 1$ there is the gapless ``XY'' phase. For $J_z>1$, the system is in the gapped antiferromagnetic N\'{e}el phase. The phase transition at $J_z=-1$ is of second order and the one at $J_z=1$ is of Berezinsky-–Kosterlitz-–Thouless type \cite{Schollwoeck2004magnetism}.
Using comparable (moderate) computer resources, the energy bound obtained from the SDP method \eqref{eq:gspRelaxed} improves substantially on the Anderson bound \eqref{AndersonBound}; see Fig.~\ref{fig:XXZ1D}. For $J_z=0$, where the model corresponds to a system of free fermions, the SDP bounds reproduce the exact ground-state energy to high precision. Employing higher-order Green's functions tends to improve bounds at larger $J_z$. The obtained short-range correlators coincide very well with the exact Bethe ansatz results \cite{Cloizeaux1966-7,Kato2003-36}.

\subsection{2D XXZ model}
\begin{figure}
	\includegraphics[width=\linewidth]{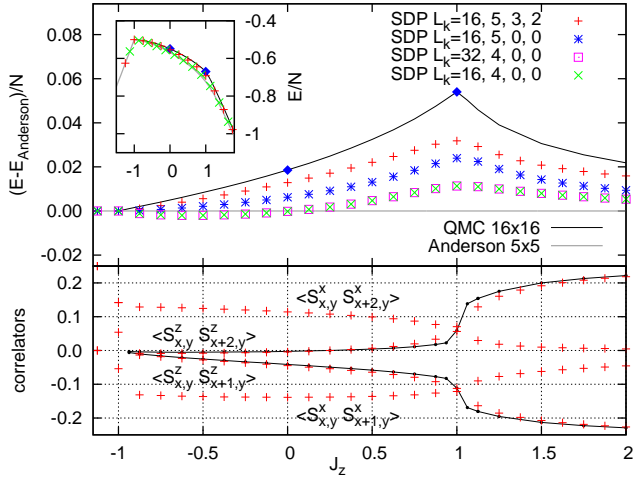}
\caption{\label{fig:Heisenberg2D} Lower energy bounds and correlators for the 2D XXZ model, complemented by QMC data calculated for a square lattice of $16\times 16$ sites with periodic boundary conditions, and inverse temperature $\beta=96$. The QMC energies for $J_z=0,1$ coincide with earlier results (\blue{$\blacklozenge$}) from Refs.~\cite{Sandvik1997-56,Sandvik1999-60}. The QMC error bars would be smaller than the line width, and another simulation with a $32\times 32$ lattice produced visually indistinguishable results. As SDP constraint subsystems $\Omega_k$ [Eq.~\eqref{eq:constraintOp}] we chose $L_k\times L_k$ squares with $K=4$ and $L_1,L_2,L_3,L_4$ as specified in the legend.}
\end{figure}
Let us now consider the spin-$1/2$ XXZ model \eqref{eq:HamXXZ} on a square lattice. With the identification $\A_i^\pdag=\S_i^{-}$, $\A_i^\dag=\S_i^{+}$, $\S^z_i=\A_i^\dag\A_i^\pdag-\frac{1}{2}$,
it maps to a model of interacting hard-core bosons obeying the algebra $\A_i^\pdag\A_j^\dag-(-1)^{\delta_{ij}}\A_i^\dag\A_j^\pdag=\delta_{ij}$. 
As displayed in Fig.~\ref{fig:Heisenberg2D}, the SDP method yields much better lower bounds to the ground-state energy than the Anderson bound.
As there is no exact solution available, we also simulated the model with QMC based on the stochastic series expansion with directed loops \cite{Syljuasen2002-66}. It is again established that the SDP method gives access to the correlation functions. 

\subsection{2D $t$--$V$ Hubbard model of spinless fermions}
\begin{figure}
	\includegraphics[width=\linewidth]{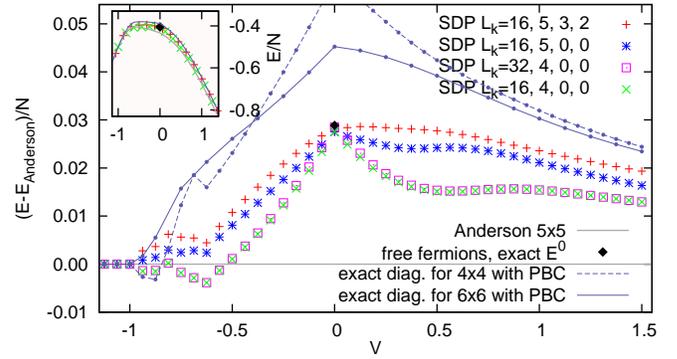}
\caption{\label{fig:FermiHubbardSpinless2D} Lower energy bounds for the 2D $t$--$V$ Hubbard model of spinless fermions on a square lattice \eqref{eq:FermiHubbardSpinless2D}, compared to exact energies for $4\times 4$ and $6\times 6$ lattices with periodic boundary conditions (PBC) as well as the thermodynamic limit at $V=0$.}
\end{figure}
Finally, Fig.~\ref{fig:FermiHubbardSpinless2D} shows results of the SDP method for the 2D $t$--$V$ Hubbard model of spinless fermions
\begin{equation}\label{eq:FermiHubbardSpinless2D}
	 \hat H = -\frac{1}{2}\sum_{\bra i,j\ket} (\A^\dag_i\A^\pdag_j + h.c.) + V \sum_{\bra i,j\ket}(\hat{n}_i-\frac{1}{2})(\hat{n}_j-\frac{1}{2})
\end{equation} 
on a square lattice.
In this case, except for $V=0$, no exact results are available. QMC is in this case hampered by the sign problem \cite{Hirsch1982-26,Takasu1986-75,Troyer2005}. It is hence inefficient and only applicable for small system sizes. The exact diagonalization results for small lattices show strong finite-size effects. The SDP method, however, is applicable just as well as for the other models, outperforms the Anderson bound, and reproduces the exact result for $V=0$. Hence, we have a completely new and promising route to easily and controlledly study frustrated magnets and fermionic systems in $D\geq2$, which theorists laboriously try to address since decades whilst being confronted with big methodological hurdles.

\section{Bipolar theorem}
Often, one is only interested in the single- and two-particle Green's functions. However, in the presented SDP approach, we also introduce higher Green's functions to improve the approximation. On the basis of the bipolar theorem, one can understand that such higher Green's functions represent slack variables.
The set of representable $(k\leq P)$-point Green's functions
\begin{equation*}
	\tilde{\R}_P:=\{\G\,|\, \exists\,\dm\in\D:\, \G^{(k)}_\sig=\Tr\dm\sig \,\,\,\forall_{k\leq P,\, \sig\in \mc{A}^k} \}
\end{equation*}
is convex, as $\G$ is linear in $\dm$ and $\D$ is convex.
Giving up on the (inessential) normalization of the Green's functions, the set $\R_P:=\{\alpha \G\,|\, \G\in \tilde{\R}_P,\,\alpha\in \RR_+ \}$ becomes a convex cone.
For a given scalar product, $\bra \cdot,\cdot\ket$, the bipolar theorem \cite{Rockafellar1970} states that $(\R_P^*)^*=\R_P$, where 
\begin{equation}
	\R_P^*:=\{\hat B\,|\,\bra B,\G\ket\geq 0\,\,\,\forall_{\G\in \R_P}\}
\end{equation} 
is the polar cone of $\R_P$.
With the choice $\bra B,\G\ket:=\sum_k\sum_{\sig\in \mc{A}^k}\G^{(k)}_\sig B_\sig\equiv \G[\hat B]$, the polar $\R_P^*$ is the convex cone of all positive-semidefinite operators from $\B_P:=\Span \bigcup_{k=0}^P\mc{A}^k$.
Due to the bipolar theorem, $\R_P$ is hence characterized by $\R_P^*$ as
\begin{equation}
	\R_P =\{\G \,|\,\G_{\sig^\dag}=\G_\sig^*\,\forall_{\sig},\, \G[\hat B]\geq 0 \,\,\,\forall_{\hat B\in \B_P,\, \hat B\succeq 0} \}.
\end{equation}
So, to obtain (or approximate) the $(k\leq P)$-point Green's functions, one needs to consider only $(k\leq P)$-point operators $\hat B\succeq 0$. In this sense, higher Green's functions are slack variables, which are only employed in order to bring the ground-state problem into the form of an SDP; see Eqs.~\eqref{eq:gspRelaxed} and \eqref{eq:SDPconstraint}. 
We showed how constraints $G[\hat B]\geq 0$ can be enforced in the SDP, for the case that $\hat B=\hC^\dag\hC$ with constraint operators $\hC$ that are polynomials of degree $\leq P/2$. However, there are also subspaces of operators $\hC$ of polynomial degree $>P/2$ such that $G[\hC^\dag\hC]$ can be evaluated with the $(k\leq P)$-point Green's functions. They can hence be taken into account without introducing higher Green's functions. A particularly simple space of such operators for a particle-number-conserving system is given by $\hC=\sum_{\vec{i}} c_\vec{i} \A_{i_1} \dots \A_{i_m}+h.c.$: For every odd $m$, $G[\hC^\dag\hC]$ can be evaluated without requiring $G^{(2m)}$.

\section{Conclusion}
We have presented a method for calculating lower bounds to the ground-state energy of condensed-matter systems and approximations to the ground-state Green's functions.
Based on certain relaxations of the ground-state problem one obtains efficiently solvable SDPs. The method can also be used for systems with particles of mixed statistics, higher spins, etc., by employing the corresponding operator algebras.
Our generic considerations on the SDP method carry over to quantum chemistry problems. An advantage in condensed-matter applications is that translation invariance and locality can be exploited to systematically balance the precision and the computation cost. Still, the idea of restricting the set of constraint operators to a physically motivated subset is also applicable to quantum chemistry problems.

\acknowledgments
We thank J.\ Eisert for calling our attention to the calculation of energy bounds using SDP, P.\ Corboz, J.\ Eisert, H.~G.\ Evertz, D.~A.\ Mazziotti, A.\ Sandvik, and M.\ Troyer for helpful discussions, A.~M.\ L\"{a}uchli for kindly providing exact diagonalization data for Fig.~\ref{fig:FermiHubbardSpinless2D}, and participants of the DPG meeting, March 2011, for useful comments.

\appendix

\section{Semidefinite programming}\label{sec:sdp}
For the determination of lower energy bounds we use semidefinite programming (SDP) \cite{Vandenberghe1996-38,Alizadeh1995-5}. The SDP algorithm iteratively approaches the solution of two different but related numerical problems, one corresponding to the lower energy bound we are looking for. Under certain circumstances, the limiting point is a guaranteed optimum of both optimization problems and provides us with a certified lower bound for the energy.

Let $Q$ and  $M_{\sigma}$ with $\sigma = 1,\dots,m$ be symmetric real matrices and $H \in \RR^m$. The first, and so called \emph{primal problem}, is to find
\begin{align}
p^* :=& \max_{X \succeq 0} \Tr{Q^T X}\\
& \text{such that }  \Tr{M_{\sigma}^T X} = H_{\sigma}\quad \forall\,  \sigma.
\end{align}
The second, and so called \emph{dual problem}, is to find
\begin{align}
d^* :=& \min_{g \in \RR^m} \sum_{\sigma = 1}^m g_\sigma H_\sigma \\
& \text{such that } \sum_{\sigma = 1}^m g_{\sigma} M_{\sigma} - Q \succeq 0.
\end{align}
The dual problem is our energy minimization problem if we let $g_{\sigma} = G_{\hat{\sigma}}$ be the Green's function, $M_{\sigma} = M_{\hat{\sigma}}$ the constraint matrices occurring in the positivity condition Eq.~\eqref{eq:SDPconstraint}, and $H$ the Hamiltonian in its vectorized representation. Let also $Q = - M_\Id$, without coefficient, because $G_\Id \equiv 1$ is fixed, as it corresponds to the norm of the state we are searching for.

We always have that $p^* \leq d^*$, because the cone of positive-semidefinite matrices is self-dual, implying $X,Y \succeq 0 \Rightarrow \Tr{X^T Y} \geq 0$ and, hence,
\begin{align}
0 & \leq \Tr{\big(\sum_{\sigma = 1}^m g_{\sigma} M_{\sigma} - Q\big)^T X} = \sum_{{\sigma} = 1}^m g_{\sigma} \Tr{M_{\sigma}^T X} - \Tr{Q^T X} \nonumber\\
& = \sum_{\sigma = 1}^m g_\sigma H_\sigma - \Tr{Q^T X}.
\end{align}
The strong duality theorem states that if the dual problem is \emph{strictly feasible}, i.e., $\exists g : \sum_{\sigma = 1}^m g_{\sigma} M_{\sigma} - Q \succ 0$ and $d^* > -\infty$, then $p^* = d^*$ and, in particular, $p^*$ and $d^*$ attain their supremum and infimum, respectively. In fact, the relaxed ground-state problems that we describe in this article are strictly feasible, as one can always construct a  mixed quantum state (corresponding to a particular vector $g$) that yields positive expectation values for all positive-semidefinite constraint operators under consideration.
Hence, there is no gap between the limiting points of the SDP and the solution is the actual infimum of the energy minimization problem.

During the optimization procedure, the program searches for data underlying an improved result for $d^*$ or $p^*$ within, or at the boundary of, convex cones. One cone is the set $\{X | X \succeq 0\}$ the other is $\{ Z | Z = \sum_{\sigma = 1}^m g_{\sigma} M_{\sigma} \succeq Q \}$. Different from, e.g., the simplex algorithm known from linear programming, which iterates along the boundary of the set of solutions, we make use of interior-point solvers, which show a superior performance in the context of SDP. These solvers consider modifications of the original optimization problem and advance along a path \emph{within} the convex cone. Given certain assumptions, using interior point methods, the SDP can be solved to any desired numerical precision within polynomial time.

\section{Explicit examples for constraints}\label{sec:constraints_k12}
Let us now exemplify the construction of the constraint matrices $M_\sig$, occurring in Eq.~\eqref{eq:SDPconstraint}, for translation-invariant particle-number-conserving fermionic ($\eta=-1$) or bosonic ($\eta=+1$) systems.
\begin{equation}
	\A_i\A_j^\dag - \eta\A_j^\dag\A_i = \delta_{ij}.
\end{equation}
In this section, the Einstein summation convention is adopted and the standard matrix basis $\Delta_{i,j}$ with
\begin{equation}
	[\Delta_{i,j}]_{i',j'}:=\delta_{ii'}\delta_{jj'}
\end{equation}
as well as the standard vector basis $\vec{\Delta}_i$ with $[\vec{\Delta}_i]_{i'}:=\delta_{ii'}$ are used.
Let us denote the 2-point Green's function elements by $G^{(2)}_{i,i'}:=G[\A_i^\pdag\A_{i'}^\dag]$, and the 4-point Green's function elements by $G^{(4)}_{ij,i'j'}:=G[\A_i^\pdag\A_j^\pdag\A_{j'}^\dag\A_{i'}^\dag]$.
The constraint matrices are constructed according to the general procedure described in Sec.~\ref{sec:Method}. For brevity, we restrict the presentation here to Green's functions $G^{(k)}$ with $k\leq 4$, but have also employed higher Green's functions in the simulations. Please note that in the quantum chemistry literature, $G^{(2)}$ is known as the ``single-particle reduced density matrix'', often denoted by $\rho$, and $G^{(4)}$ is known as the ``two-particle reduced density matrix'', often denoted by $\Gamma$. For the following reasons we think that it is in this case preferable to adhere to the condensed-matter-theory convention of calling the quantities $G^{(k)}$ \emph{(equal-time) Green's functions} which is a synonym for correlation functions of ladder operators \cite{Fetter1971,Negele1988}:
(a) The $G^{(k)}$, as matrices, are not normalized to 1 and naturally occur in their unnormalized form in the SDP. (b) A normalization is neither useful nor easily possible, unless one considers the full Green's function of a finite particle-number-conserving system. (c) For some models, one also needs to consider $G^{(k)}$ for odd $k$ which can not be interpreted as a kind of density matrix.

Now, let us first consider only one-point constraint operators, i.e., $\hC$ from the set $\Csq=\Span \mc{A}^0\cup\mc{A}^1_{\Omega_1}$ with $\mc{A}^k_\Omega$ as in Eq.~\eqref{eq:opBasis}, i.e., $\hC=C_{\none,\none} +C_{i,\none}\A_i+C_{\none,i}\A^\dag_i$. The evaluation of the corresponding positive-semidefinite observables $\hC^\dag\hC$ with respect to the Green's function reads
\begin{equation*}
	G[\hC^\dag\hC] = |C_{\none,\none}|^2 + C^*_{\none,i} G[\A_i^\pdag\A_j^\dag] C_{\none,j} + C^*_{i,\none} G[\A_i^\dag \A_j^\pdag] C_{j,\none}
\end{equation*}
and $G[\hC^\dag\hC]\geq 0$ $\forall_{\vec{C}}$ is hence equivalent to
\begin{subequations}
\begin{gather}
	G[\A_i^\pdag\A_j^\dag]\Delta_{i,j} = G^{(2)}\succeq 0,\\
	G[\A_i^\dag \A_j^\pdag]\Delta_{i,j} = \eta G^{(2)\T}-\eta\id\succeq 0.
\end{gather}
\end{subequations}

Now we consider two-point constraint operators, i.e., $\hC$ from the set $\Csq=\Span \mc{A}^0\cup\mc{A}^1_{\Omega_1}\cup\mc{A}^2_{\Omega_2}$ with some appropriate choice for the subsystems $\Omega_k$. The evaluation of $\hC^\dag\hC$ with respect to the Green's function reads with $\hC=C_{\none,\none} +C_{i,\none}\A_i+C_{\none,i}\A^\dag_i+C_{ij,\none}\A_i\A_j+C_{i,j}\A^\pdag_i\A^\dag_j+C_{\none,ij}\A_j^\dag\A_i^\dag$
\begin{align*}
	G[\hC^\dag\hC] &= |C_{\none,\none}|^2 + C^*_{\none,i} G[\A_i^\pdag\A_{i'}^\dag] C_{\none,i'}\\
	 &+ C^*_{i,\none} G[\A_i^\dag \A_{i'}^\pdag] C_{i',\none}\\
	 &+ C^*_{\none,ij} G[\A_{i}^\pdag\A_{j}^\pdag\A_{j'}^\dag\A_{i'}^\dag] C_{\none,i'j'}\\
	 &+ C^*_{ij,\none} G[\A_{j}^\dag\A_{i}^\dag\A_{i'}^\pdag\A_{j'}^\pdag] C_{i'j',\none}\\
	 &+ C^*_{i,j} G[\A_{j}^\pdag\A_{i}^\dag\A_{i'}^\pdag\A_{j'}^\dag] C_{i',j'}\\
	 &+ (C_{\none,\none}^* G[\A_{i'}^\pdag\A_{j'}^\dag] C_{i',j'} + c.c.)
\end{align*}
and $G[\hC^\dag\hC]\geq 0$ $\forall_{\vec{C}}$ is hence equivalent to
\begin{subequations}
\begin{gather}
	G[\A_i\A_{i'}^\dag] \Delta_{i,i'} = G^{(2)}\succeq 0,\\
	G[\A_{i'}^\dag \A_{i}^\pdag] \Delta_{i,i'} = \eta G^{(2)}-\eta\id\succeq 0,\\
	G[\A_{i}^\pdag\A_{j}^\pdag\A_{j'}^\dag\A_{i'}^\dag] \Delta_{ij,i'j'} =  G^{(4)}\succeq 0,\\
\begin{split}
G[\A_{j'}^\dag&\A_{i'}^\dag\A_{i}^\pdag\A_{j}^\pdag] \Delta_{ij,i'j'}\\
	 =& \big[G^{(4)}_{ij,i'j'} -\delta_{jj'}G^{(2)}_{i,i'} -\delta_{ii'}G^{(2)}_{j,j'} \\
	 & -\eta\delta_{i'j}G^{(2)}_{i,j'} -\eta\delta_{ij'}G^{(2)}_{j,i'} \\
	 & +\delta_{ii'}\delta_{jj'} +\eta\delta_{i'j}\delta_{ij'}  \big]\Delta_{ij,i'j'} \succeq 0,
\end{split}\\\label{eq:constraint_k2}
	\tmat{1&G[\A_{i'}^\pdag\A_{j'}^\dag]\vec{\Delta}^\T_{i'j'}\\G[\A_{j}^\pdag\A_{i}^\dag]\vec{\Delta}_{ij}&G[\A_{j}^\pdag\A_{i}^\dag\A_{i'}^\pdag\A_{j'}^\dag]\Delta_{ij,i'j'}}\succeq 0.
\end{gather}
\end{subequations}
These matrix expressions are linear in $G^{(2)}$ and $G^{(4)}$.

\section{Anderson bounds}\label{sec:Anderson}
Anderson bounds \cite{Anderson1951-83,Wittmann1993-48} are determined by splitting the Hamiltonian $\hat H$ into $M$ terms $\hat H_m$ for which the ground-state energies are accessible, for example, by choosing the terms $\hat H_m$ such that their supports are small enough to allow for an exact diagonalization of each term.
\begin{equation}\textstyle
	\hat{H}=\sum_{m=1}^M\hat{H}_m
\end{equation}
The computation cost for determining the ground-state energies of the operators $\hat{H}_m$ scales exponentially in the size of their spatial supports. One can exploit the sparseness of the Hamiltonians and access, with state of the art computer resources, subsystem sizes of up to about 25 spins-$1/2$. Larger subsystems are difficult to address, as the $\hat{H}_m$ are not translation-invariant. With diagonal representations $\hat{H}_m=\sum_{m,n}E^n_{m}|m,n\ket\bra m,n|$ and the lowest energy eigenstates $|m,0\ket$, one has
\begin{gather}\textstyle
	\hat{H}_m\succeq \sum_{m,n} E^0_m|m,n\ket\bra m,n|=E^0_m\cdot\id\\\textstyle
	\Rightarrow
	\hat{H}\succeq \sum_m E^0_m\cdot \id,
\end{gather}
i.e., $\sum_m E^0_m$ is a lower bound to the ground-state energy of the Hamiltonian $\hat H$.

\section{Related work}\label{sec:Literature}
In Refs.~\cite{Hammond2006-73,Nakata2008-128,Shenvi2010-105}, the SDP method employing the full single-particle and two-particle Green's functions, as developed in the context of quantum chemistry, has been applied to the 1D Hubbard model with up to 14 sites.
Verstichel \emph{et al.} \cite{Verstichel2010-132} as well as Shenvi and Izmaylov \cite{Shenvi2010-105} improve upon the original SDP approach, using the full single-particle and two-particle Green's functions, by adding further linear inequalities. Those inequalities are of the form $G[\hat X]\geq X_0$ for specific operators $\hat X$ which are supported in certain subsystems. In Ref.~\cite{Shenvi2010-105}, the size of the subsystems is chosen such that $\hat X$ can be diagonalized exactly and $X_0$ is the corresponding lowest eigenvalue. In Ref.~\cite{Verstichel2010-132} the dissociation of molecules is studied. To this purpose $\hat X$ is chosen to be a restriction of the full Hamiltonian to a single atom. $X_0$ is chosen as a lower bound to the spectrum of $\hat X$, determined by solution of a separate SDP. Corresponding additional linear constraints could also be useful for further improving the results of the SDP method for condensed-matter problems, as developed in this work.
In Ref.~\cite{Pironio2010-20}, Pironio, Navascu\'{e}s, and Ac\'{\i}n describe convergent relaxations of polynomial optimization problems with non-commuting variables. The method generalizes Lasserre's algorithm for commuting variables \cite{Lasserre2001-11}, is applicable to models in second quantization, and works in a representation free way. To this purpose, the commutation relations of the ladder operators are implemented as a set of constraints to the SDP. It is not obvious how to address, within this approach, the thermodynamic limit or how to implement translation invariance.

\emph{Note added:} Shortly after the online publication of this work, T.~Baumgratz and M.~B.\ Plenio have presented arxiv:1106.5275v1 in which they calculate lower energy bounds to many-particle systems on finite lattices. It is based on an SDP approach employing the full single-particle and two-particle Green's functions and a modified algorithm for the solution of semidefinite programs.

\bibliographystyle{prsty} 

\end{document}